\documentclass[balance,upint,subscriptcorrection,varvw,mathalfa=cal=boondoxo,spanish,french,vietnamese,russian,greek,pdf-a,colorlinks]{asmeconf}

\hypersetup{%
	pdfauthor={John H. Lienhard},									  
	pdftitle={ASME Conference Paper LaTeX Template},                  
	pdfkeywords={ASME conference paper, LaTeX template, BibTeX style},
	pdfsubject = {Describes the asmeconf LaTeX template},			  
	pdflicenseurl={https://ctan.org/pkg/asmeconf},
}

\begin{document}


\ConfName{Proceedings of the OMAE 2024\linebreak 43rd International Conference on Ocean, Offshore \& Arctic Engineering}
\ConfAcronym{OMAE2024}
\ConfDate{June 9–14, 2024} 
\ConfCity{Singapore} 
\PaperNo{OMAE2024-125991}


\title{Modeling of Hydroacoustic Noise from Marine Propellers with Tip Vortex Cavitation} 
 
%
%
%

\SetAuthors{%
	Zhi Cheng\affil{1}\CorrespondingAuthor{vamoschengzhi@gmail.com}, 
    Suraj Kashyap\affil{1},
	Brendan Smoker\affil{2},  
	Giorgio Burella\affil{2},  
	Rajeev Jaiman\affil{1}\CorrespondingAuthor{rjaiman@mech.ubc.ca}
	}

\SetAffiliation{1}{Mechanical Engineering, The University of British Columbia, 2329 West Mall, Vancouver, British Columbia, V6T 1Z4, CANADA. }
\SetAffiliation{2}{Robert Allan Ltd. - Naval Architects and Marine Engineers, 250 Howe St \#400, Vancouver, British Columbia, V6C 3R8, CANADA.}

{
\maketitle
}
\versionfootnote{Documentation for \texttt{asmeconf.cls}: Version~\versionno, \today.}


\keywords{Marine Propeller, Large-Eddy Simulation, Turbulent Wake, Tip Vortex Cavitation, Hydroacoustics}


\begin{abstract}
Due to growing marine ecological concerns, there is an acute industrial need to significantly reduce the underwater radiated noise (URN) from marine propellers during ship operations. In this regard, high-fidelity fluid flow and hydroacoustic models are required for understanding the propeller noise generation and propagation in the ocean environment. Using high-fidelity CFD modeling, the present work aims to study the cavitating turbulent flow of a full-scale marine propeller and explore the physical mechanism underpinning the underwater radiated noise. We employ the standard dynamic large eddy simulation for the turbulent wake flow and the Schnerr-Sauer cavitation model, while the  Ffowcs-Williams-Hawkings acoustic analogy is considered for the hydroacoustic modeling. For the current investigation, we consider a well-known Potsdam Propeller Test Case to analyze the turbulent cavitating flow and the associated hydroacoustic emissions. To begin, the modeling framework is validated using the available experimental data, and distinctive double-helical tip vortex cavitation and its qualitative patterns along the vortex trajectory are captured.
In comparison to the non-cavitating condition, the pressure distribution on the propeller surface is more disordered for the cavitating condition, which is further reflected by a relatively stronger power of both low-frequency tonal peaks and high-frequency broadband components in the spectrum of thrust generation. Specifically, the generation of cavitation leads to the enhancement of the monopole noise source and the breakdown of cavitation bubbles as well as vortex structures in the turbulent wake. Furthermore,  the tonal noise with the frequency corresponding to the harmonics of blade passing frequency is also enhanced. Generally speaking, the generation of cavitation structures enhances the hydroacoustics energy of URN at all orientations, especially in the downstream direction with sound pressure level increasing up to 20 dB.

\end{abstract}


%










\section{Introduction}
The generation and propagation of underwater radiated noise (URN) generated from marine propellers \citep{SMITH2022112863} is one long-term industrial and environmental issue, that negatively affects the safety and health of ship crew, passengers as well as marine mammals \citep{Savas2021, Asnaghi2020, A-Man2023}. 
URN substantially includes machinery noise, propeller noise, and hydrodynamic flow noise. The hydrodynamic flow noise, caused by the mutual motion of the ship hull and the surrounding water, is generally weak and also usually covered by other noises \cite{Arveson2000}. 
Regarding the hydrodynamic noise induced by operating propellers, it is usually characterized by tonal noise at different blade pass frequencies, as well as broadband noise originating from various wake vortexes, cavitation structures, and bursting bubbles. 
In addition, modern propeller designs often purposefully allow certain non-erosive cavitation to occur to achieve higher efficiencies, which will exacerbate the energy of radiated noise \cite{jmse9070778}. 
The dramatic noise increases caught by propeller cavitation will cover mechanical noise and dominate URN propagation \cite{seol2002prediction}.
Therefore, it is extremely significant to predict and analyze the noise-radiating behaviors accompanying the operation of cavitating propellers.

The rapid development of computational resources makes it possible to numerically predict the turbulence field, cavitation development, and noise propagation of the full-scale marine propeller. Hybrid methodology of acoustics calculation reduces the computational effort while guaranteeing predictive accuracy and, as a consequence, is more amenable to the calculation of noise propagation, especially in applications addressing engineering and real-world industrial problems. For the prediction of sound radiation in the hybrid method, the commonly used methodology is based on the Ffowcs-Williams-Hawkings (FW-H) equations \cite{VariantsoftheFfowcsWilliams, OntheuseoftheFfowcsWilliams, FILIOS20071497}.
Posa {\it et al.} \cite{posa_JFM_2022} investigated the flow dynamics and acoustics propagation of marine propellers via the database (consisting of 840 million points) generated by large-eddy simulations (LES) and FW-H method. 
Cavitation is not considered in this work. Posa {\it et al.} \cite{posa_JFM_2022} observed that the tonal components of acoustic signature mainly appear in the near ﬁeld and dissipate rapidly when extending to the far field. The energy of tonal noise is transferred into that of broadband components because the noise sources underpinning tonal components develop instability at a faster rate.
Similar work was also conducted by Ahmed \cite{Shakeel2020}. 

Posa {\it et al.} \cite{Posa2022} further considered the interaction of the upstream propeller and downstream hydrofoil and state that dominant noise sources are located at the leading edge of the hydrofoil because of the impingement by the propeller wake.
Yangzhou {\it et al.} \cite{yangzhou_wu_ma_huang_2023} investigated 
a similar configuration and proposed two different noise source-identifying approaches.
Compared to the open propellers introduced above, the nozzle surrounding rotors in the ducted condition will have an impact on both the flow dynamics and acoustics generation. As introduced in the works of Zhang \& Jaiman \cite{ZHANG2019202}, the flow through the tip-leakage ( between the nozzle and the blade tip) will generate the tip leakage vortex, which will interact with the trailing edge wake and tip vortices shed from the blades. This phenomenon will pose some potential implications for the noise sources generated during propeller operation.
Hieke {\it et al.}\cite{HIEKE2022112131} and Chen {\it et al.} \cite{jmse10030378} both experimentally and numerically investigated the noise characteristics of the ducted marine propeller, but did not involve cavitation modeling in the calculations. 

In terms of the potential effects of cavitation generation on flow dynamics and noise sources for full propellers, Potsdam Propeller Test Case (PPTC) of smp workshop \cite{PPTC_propeller} published one propeller model (termed as VP 1304) and conducted a series of measurements on the cavitating formulation. There are some subsequent computational works such as Geese \& Kimmerl \cite{Geese2022}, Viitanen {et al.} \cite{VIITANEN2022112596}, Cianferra \& Petronio \cite{Marta2019}, in which their simulation results were compared with the experimental data.
For the numerical works on the full propeller cavitation, 
Kimmerl {\it et al.} \cite{jmse9070778} included the cavitation processing in consideration and also investigated the effect of the shock of propeller wake on the hull surface.
The cavitation process is also considered in the computational work of Lidtke {\it et al.} \cite{LIDTKE2022111176} and the difference in noise characteristics between the ducted and non-ducted conditions is compared. The computational results implied that the addition of a nozzle effectively suppresses the tonal components of the noise spectrum. However, this work does not obtain a sufficient and accurate prediction of cavitation structures while predicting hydroacoustic emission.

In view of the aforementioned literature, it is important to investigate the hydro-acoustic noise generated by a three-dimensional turbulent and cavitating wake behind the marine propellers and compare the discrepancies between cavitating and non-cavitating configurations. 
In the present work, two scenarios, in detail, VP 1304 propellers without cavitation model and with cavitation model are considered with cross-comparison with respect to the propeller performance, the cavitation structures, and the noise generation. 
More specifically, the emphasis will be placed on the surface pressure distribution, the specific pattern of cavitation, and the detailed variation/transfer of the shedding tip vortex.
Moreover, this study further explores their implication on noise characteristics and power requirements.

The paper is structured as follows:
Section 2 details the numerical and analytical methods listed above.
Section~3 describes the present concerned problem.
The accuracy of the implemented models used herein is also validated in Section~3.
In Section 4, the flow dynamics of the propeller with cavitating and non-cavitating conditions are analyzed. Section 5 further explores the hydroacoustics performance.
In Section 6, the key results of this study are summarized.

\section{Methodology}

\subsection{Fluid dynamics and solid body motion}

The fluid domain consists of liquid and vapor phases which are assumed to exist as a continuous homogeneous mixture. The phase indicator $\alpha _l$($\boldsymbol{x}$,$t$) represents the phase fraction of the liquid phase in the fluid mixture, where $\boldsymbol{x}$ and $t$ are spatial and temporal coordinates. The density $\rho$ and dynamic viscosity $\mu$ of the fluid are obtained as a linear weighted combination of the liquid and vapor phases:
	\begin{equation}\label{}
\rho =\rho _l\alpha _l+\rho _v \left(1 - \alpha _l \right) ,
	\end{equation}
	\begin{equation}\label{}
\mu =\mu _l\alpha _l+\mu _v \left(1 - \alpha _l \right) ,
	\end{equation}
where $\rho_l$ and $\rho_v$ are the densities of the pure liquid and vapor phases respectively, and $\mu_l$ and $\mu_v$ are the corresponding dynamic viscosities. 


We model the unsteady, incompressible Newtonian flow of the fluid in the present work. The standard dynamic large eddy simulation (LES) model \cite{LESmodel} is applied here, and the unsteady Navier–Stokes equations (subjected to Favre filtering operation) describing the conservation of momentum and mass are expressed as
\begin{equation}\label{LES_eq1}
\frac{\partial \tilde{u}_j }{\partial x_j}=\left( \frac{1}{\rho_l}-\frac{1}{\rho_v} \right)\dot{m},
\end{equation}
\begin{equation}\label{LES_eq2}
\frac{\partial \left( \rho \tilde{u}_i \right)}{\partial t}+\frac{\partial \left( \rho \tilde{u}_i\tilde{u}_j \right)}{\partial x_j}=-\frac{\partial \tilde{p}}{\partial x_i}+\frac{\partial}{\partial x_i}\left( \mu \frac{\partial \tilde{u}_i}{\partial x_j} \right) -\frac{\partial \tau _{ij}}{\partial x_j},
\end{equation}
where the over-bar `$\sim$' denote filtered quantities. \(x_i\) is the $i$-th component of a Cartesian coordinate vector $\boldsymbol{x}$ with $i$, $j \equiv 1$, 2, 3 corresponding to the $x$-, $y$- and $z$-directions, respectively; \(p\) is the pressure; and, \(u_i\) represents the $i$-th component of fluid velocity. Although the pure liquid and pure vapour phases are considered incompressible, the density of the mixture varies with the volume fraction of the immiscible phases, and thus the local divergence of velocity is non-zero as seen in the continuity equation (Eq.~\ref{LES_eq1}). $\dot{m}$ is the mass transfer rate between the phases due to cavitation and has been elaborated further in Section \ref{Sec_cav_model}.
$\tau _{ij}$ in Eq.~\ref{LES_eq2} are the Sub-Grid Scale (SGS) stresses defined as
\begin{equation}\label{}
\tau _{ij}=\widetilde{u_iu_j}-\tilde{u}_i\tilde{u}_j,
\end{equation}
the Boussinesq hypothesis is used to model the SGS stresses
\begin{equation}\label{}
\tau _{ij}-\frac{1}{3}\tau _{kk}\delta _{ij}=-2\mu _t\widetilde{S_{ij}},
\end{equation}
where $\tau_{kk}$ is the isotropic part of the SGS stresses and $\mu_t$ is the SGS turbulence viscosity. $\widetilde{S_{ij}}$ is the rate-of-strain tensor for the resolved scale defined by
\begin{equation}\label{}
\widetilde{S_{ij}}=\frac{1}{2}\left( \frac{\partial \tilde{u}_i}{\partial x_j}+\frac{\partial \tilde{u}_j}{\partial x_i} \right) .
\end{equation}

\subsection{Cavitation modelling}\label{Sec_cav_model}

We model the effect of cavitation using a transport equation for the liquid phase fraction $\alpha_l$
\begin{equation}\label{}
\frac{\partial \left( \rho _l\alpha _l \right)}{\partial t}+\frac{\partial \left( \rho _l\alpha _lu_i \right)}{\partial x_i}= \dot{m} =\dot{m}_c+\dot{m}_v ,
\end{equation}
The source terms $\dot{m}_c$ and $\dot{m}_v$ model the mass transfer rates during the condensation and vaporization processes respectively as a result of cavitation. $\dot{m}_c$ and $\dot{m}_v$ in the present work are based on the original work by Schnerr and Sauer (Schnerr-Sauer cavitation model) \citep{sauer2001development}. The Schnerr-Sauer cavitation model is derived from a simplification of the Rayleigh-Plesset equation for spherical bubble dynamics and has been used extensively in literature for a variety of cavitating flow configurations \citep{asnaghi2017improvement, kashyap2023unsteady}. $\dot{m}_c$ and $\dot{m}_v$ are thus expressed as
\begin{equation}\label{eqm1}
\dot{m}_c=C_c\alpha _l\left( 1-\alpha _l \right) \frac{3\rho _v\rho _l}{\rho R_B}\sqrt{\frac{2}{3\rho _l\lvert p-p_v \rvert}}\text{max}\left( p-p_v,0 \right),
\end{equation}

\begin{equation}\label{eqm2}
\dot{m}_v=C_v\alpha _l\left( 1 + \alpha_{Nuc} -\alpha _l \right) \frac{3\rho _v\rho _l}{\rho R_B}\sqrt{\frac{2}{3\rho _l \lvert p-p_v \rvert}} \text{min}\left( p-p_v,0 \right),
\end{equation}
where $C_c$ and $C_v$ are the condensation and vaporization model coefficients respectively, and $p_v$ is the vapor pressure of the fluid at saturation conditions. $\alpha_{Nuc}$ and $R_B$ are the volume fraction and radius of the bubble nuclei in the fluid and are given as
\begin{equation}\label{alphaNuc}
\alpha _{Nuc}=\left( n_0\pi\frac{d_{Nuc}^3}{6}  \right) /\left( 1+n_0\pi\frac{d_{Nuc}^3}{6} \right) ,
\end{equation}

\begin{equation}\label{Rb}
R_B=\sqrt[3]{\frac{3}{4 \pi n_0} \frac{1+\alpha_{Nuc}-\alpha_l}{\alpha_l}} ,
\end{equation}
where $n_0$ is the number of nuclei per unit volume and $d_{Nuc}$ is the corresponding diameter of the nuclei.  
$C_c$, $C_v$, $\alpha_{Nuc}$ and $d_{Nuc}$ are user-inputs to the cavitation model. In the present study, these inputs have been set to $C_c = 1.0$, $C_v = 1.0$, $\alpha_{Nuc} = 1.6\times 10^{13}$ and $d_{Nuc} = 6\times 10^{-5}m$.

\subsection{Acoustic modelling}
The FW-H method could be used to predict the noise generated by the solid surface with arbitrary motion. Accompanying FW-H method gradually regarded as a theoretical basis for industrial sound prediction, many methods for solving the FW-H equations have been proposed.
The noise sources of the wave equation in the FW-H method contain three terms, namely, quadrupole source (noise due to turbulent shear stress), dipole source, and monopole source \cite{CHENG2023652}.

The turbulence term (quadrupole source) requires volume integration in the FW-H equations, which will make it challenging to determine the integration area and to integrate in the flow surrounding the structure \cite{farassat1988uses}. The FW-H method with a permeable control surface is an improved acoustic radiation model that builds permeable surfaces and adds some reasonable assumptions 
so that the combination of loading and thickness terms on permeable surfaces will cover the sources brought by quadrupole terms within the control surface \cite{wang2006computational, farassat2007derivation}, and the detailed information refers to our previous work \cite{CHENG2023652}.
This study applies the Farassat 1A formulas \cite{farassat1981linear, farassat1975theory, farassat1998acoustic} proposed by Farassat {\it et al.} to solve FW-H equations. 
For the noise generated by the moving source, the sound pressure \(p^{'}\) at the monitor location \(\boldsymbol{x}\) and recording time \(t\) in Farassat's formulation 1A is \cite{CHENG2023652}:
\begin{equation}\label{F1Aeq1}
p^{'}\left( \boldsymbol{x},t \right) =p_{T}^{'}\left( \boldsymbol{x},t \right) +p_{L}^{'}\left( \boldsymbol{x},t \right)
\end{equation}
where the index \(T\) and \(L\) correspond to the thickness and loading terms, which are expressed as:

\begin{equation}\label{thicknessnoiseeq}
\begin{aligned}
4\pi p_{T}^{'}\left( \boldsymbol{x},t \right) =\int_{f=0}{\left[ \frac{\dot{Q}_n+Q_{\dot{n}}}{r\left( 1-M_r \right) ^2} \right]}_{ret}dS
\\
\,\,           
+\int_{f=0}{\left[ \frac{Q_n\left( r\dot{M}_r+c_0\left( M_r-M^2 \right) \right)}{r^2\left( 1-M_r \right) ^3} \right]}_{ret}dS,
\end{aligned}
\end{equation}

\begin{equation}\label{loadingnoiseeq}
\begin{aligned}
4\pi p_{L}^{'}\left( \boldsymbol{x},t \right) =\frac{1}{c_0}\int_{f=0}{\left[ \frac{\dot{L}_r}{r\left( 1-M_r \right) ^2} \right]}_{ret}dS
\\
\,\,                +\int_{f=0}{\left[ \frac{L_r-L_M}{r^2\left( 1-M_r \right) ^2} \right]}_{ret}dS
\\
\,\,                +\frac{1}{c_0}\int_{f=0}{\left[ \frac{L_r\left( r\dot{M}_r+c_0\left( M_r-M^2 \right) \right)}{r^2\left( 1-M_r \right) ^3} \right]}_{ret}dS.
\end{aligned}
\end{equation}
Here, \(M_i=v_i/c_0\) with $i=1,2,3$ are the components of the Mach vector where $c_0$ is the speed of sound; $M$ is the length (magnitude) of the Mach vector, and $\cdot$ above a parameter represents the time derivative. 
Other terms in the above equation are represented as:

\begin{equation}\label{F1Aeq4}
\begin{aligned}
\begin{array}{c}
	M_r=M_i\hat{r}_i,   \dot{M}_r=\frac{\partial M_i}{\partial \tau}\hat{r}_i,\\
	Q_n=Q_i\hat{n}_i,    \dot{Q}_n=\frac{\partial Q_i}{\partial \tau}\hat{n}_i,    Q_{\dot{n}}=Q_i\frac{\partial \hat{n}_i}{\partial \tau},\\
	L_i=L_{ij}\hat{n}_i,     \dot{L}_r=\frac{\partial L_i}{\partial \tau}\hat{r}_i,     L_r=L_i\hat{r}_i,     L_M=L_iM_i.\\
\end{array}
\end{aligned}
\end{equation}
in which \(Q_n\) and \(L_i\) are defined as
\begin{equation}\label{F1Aeq5}
\begin{aligned}
\begin{array}{c}
Q_n=Q_n\hat{n}_i=\left[ \rho _0v_i+\rho \left( u_i-v_i \right) \right] \hat{n}_i,
\\
L_i=L_{ij}\hat{n}_i=\left[ P_{ij}+\rho u_i\left( u_j-v_j \right) \right] \hat{n}_j,
\end{array}
\end{aligned}
\end{equation}
here, \(v_i\) is the surface ($f$) moving velocities, and the summation convention is used for repeated index. The moving surface is described by \(f\left( \boldsymbol{x},t \right) =0\) such that \(\hat{n}=\nabla f\) is the unit outward normal to the surface. \(P_{ij}=\left( p-p_0 \right) \delta _{ij}-\tau _{ij}\) is the compressible stress tensor, and \(\delta _{ij}\) is the Kronecker delta. In general, the viscous term in \(P_{ij}\) could be seen as negligible. Then, the stress tensor is \(P_{ij}=\left( p-p_0 \right) \delta _{ij}^a\). Additionally, in the region away from the source area, the perturbation of density is also small, and \(c^2\left( \rho -\rho _0 \right)\) is replaced by the sound pressure \(p^{'}\).
The index \(ret\) represents the integrand at the time of emission:

\begin{equation}\label{F1Aeq6}
g=\tau _{ret}-t+\frac{r}{c_0},
\end{equation}
in which, \(r=\lvert \boldsymbol{x}-\boldsymbol{y}\left( \tau _{ret} \right) \rvert\) represents the space distance from the sound source to the monitor location.
Epikhin \textit{et al.} \cite{EPIKHIN2015150} implemented the Farassat 1A formulas in OpenFOAM \cite{openfoamv2006} and we have developed a customized solver for the present study. 

\begin{figure}[h]
  \centering
  \begin{subfigure}[b]{0.8\linewidth}
    \includegraphics[width=\linewidth]{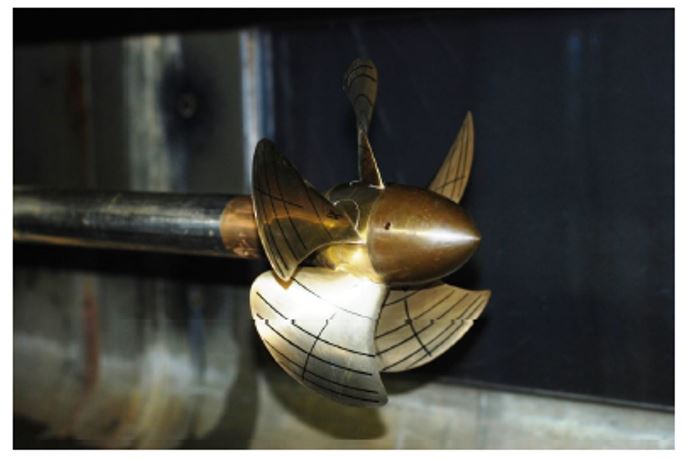}
\caption{}
\label{exp_config}
  \end{subfigure}
  
  \begin{subfigure}[b]{1.0\linewidth}
    \includegraphics[width=\linewidth]{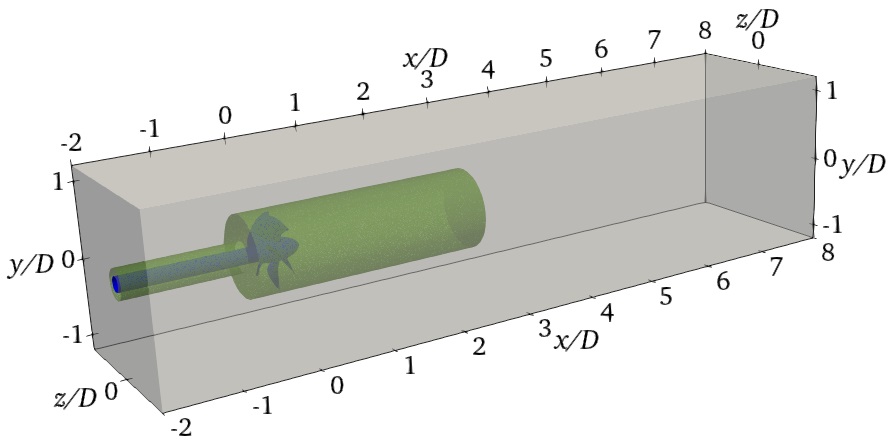}
    \caption{}
    \label{simu_config}
  \end{subfigure}
  \caption{(a) PPTC experimental configuration \cite{PPTC_propeller}, (b) 3-D computational domain with propeller (accompanied by rod) surface and sliding interface annotated by blue and green colors, respectively. }
  \label{case_config}
\end{figure}

\begin{figure}[h]
\centering
\includegraphics[width=1.0\linewidth]{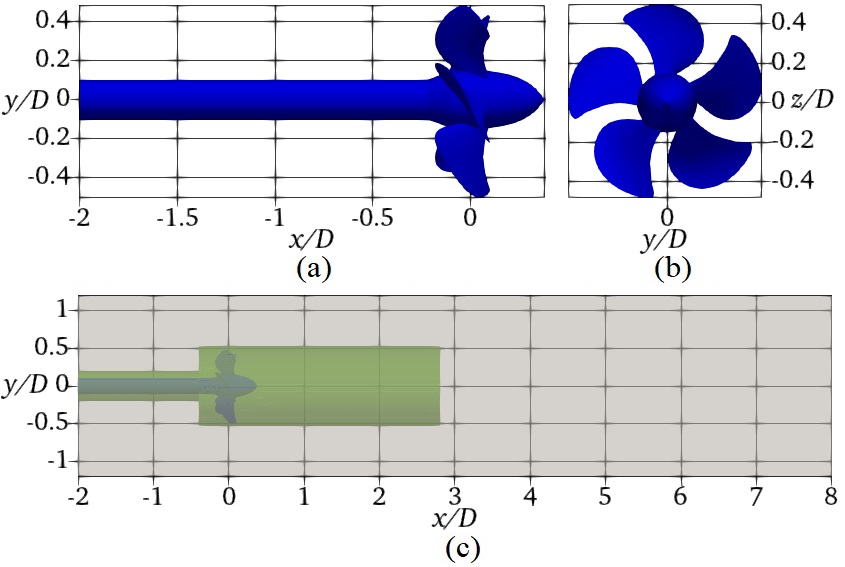}
\caption{Diagram of geometry and computational domain.}\label{configuration}
\end{figure}

\section{Description of the problem}
\label{Problemdefinition}

\begin{figure*}
  \centering
  \begin{subfigure}[b]{0.7\linewidth}
    \includegraphics[width=\linewidth]{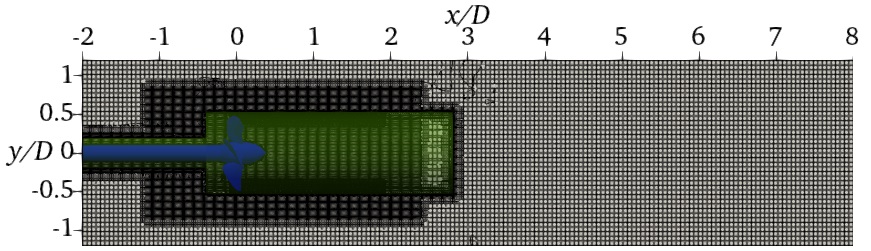}
\caption{}
\label{whole_mesh}
  \end{subfigure}
  
  \begin{subfigure}[b]{0.8\linewidth}
    \includegraphics[width=\linewidth]{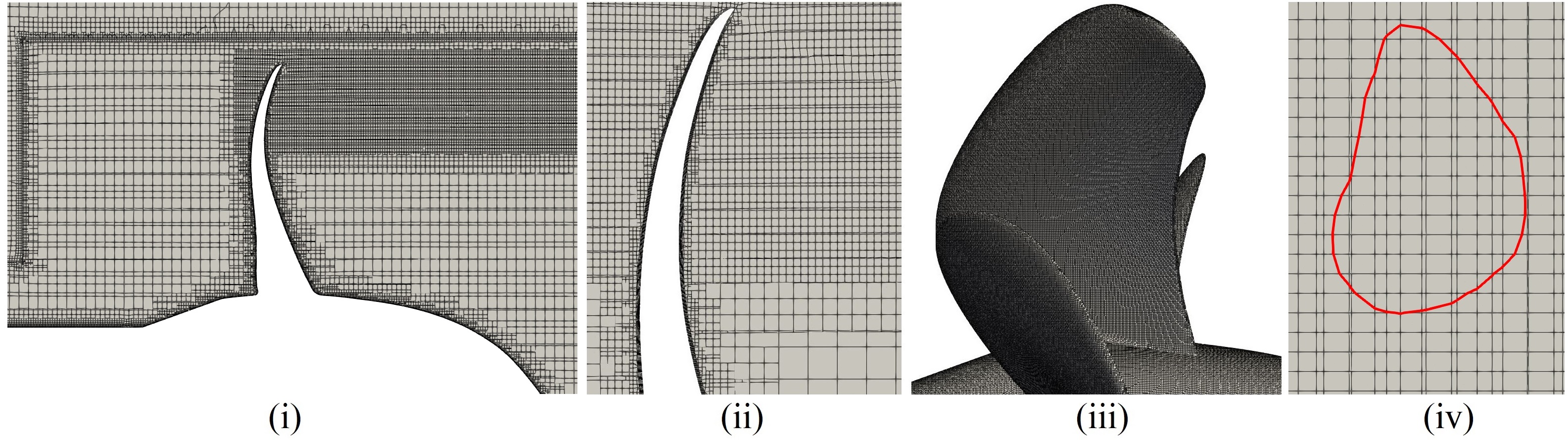}
    \caption{}
    \label{local_mesh}
  \end{subfigure}
  \caption{Mesh setup (mesh 3) used in the present simulation.
(a) Domain showing the overall mesh in one streamwise orthogonal $x-y$ plane (which is consistent with the mesh in the other streamwise orthogonal $x-z$ plane), and (b) expanded view of the immediate vicinity of the propeller surface and inplane mesh distribution (last panel iv) at $x/D$ = 0.25, the rot line represents the cavitation contour with a volume fraction of 0.5.}
  \label{mesh1and2}
\end{figure*}


Following the works of Geese \& Kimmerl \cite{Geese2022} and Cianferra \& Petronio \cite{Marta2019}, we choose the standard Potsdam Propeller Test Case (PPTC) for which experimental data was made available during the SMP workshop \cite{PPTC_propeller}. We select a particular case of the five-bladed controllable pitch marine propeller VP1304. The salient features of the propeller geometry and operating conditions are summarized in Table \ref{Propeller_pro_ope}. 

Figures \ref{case_config}a and \ref{case_config}b exhibit the experimental propeller geometry and the corresponding numerical configuration. The propeller is mounted on a cylindrical shaft and submerged in water. 
As indicated in Figure \ref{configuration}, the position of the propeller hub center is at the centerline of both transverse (cross-stream) directions (i.e., $z=y=$ 0), and located at a distance 2$D$ downstream of the inlet boundary in the $x-$direction. The streamwise ($x-$) length and two cross-stream ($y-, z-$) lengths of the computational domain are 10$D$, 2.4$D$, and 2.4$D$.
A Dirichlet boundary condition was prescribed for the incident flow velocity $\vec u = (U_0, 0, 0)$ on the inlet face (i.e., single upstream patch) in Fig. \ref{case_config}b.
A Neumann boundary condition is imposed on the velocity at the outflow (outlet) boundaries, i.e., the single downstream patch of the domain, symmetrical boundary conditions are applied to four side patches to avoid blockage effects.


\subsection{Mesh generation and model validation}
Spatial discretization of the resulting highly complex fluid domain requires careful consideration. For efficient mesh generation we use the SnappyHexMesh utility implemented in OpenFOAM. SnappyHexMesh generates high quality hexahedral or split-hexahedral meshes from triangulated surface geometries. This will enable template-based shape optimization studies which is of interest to the authors in future work.
For mesh 3 applied in this paper, it takes about 4.5 hours with 4$\times$40 G computing resources to generate the mesh with cell numbers of about 40.4 million.

\begin{center}
\begin{table}[!h]
\caption{Geometric parameters of VP1304 propeller and relevant operating conditions.}
\centering
\begin{tabular}{  c  c  c  c  c  c  c  c  c  c  }
\hline
  Properties & Value  \\
\hline 
 Pitch ratio at $r/R$ = 0.7 & 1.635   \\ 
 Chord at $r/R$ = 0.7 & 0.417 \\ 
 Skew ($^{\circ}$) & 18.837 \\ 
 Hub ratio & 0.300 \\
 Number of blades & 5 \\
 Propeller diameter $D$ (m) & 0.25   \\
 Advance coefficient $J$ & 1.019 \\ 
 Rotational speed $n$ (1/s) & 25.0 \\
 Cavitation number $\sigma _n$ & 2.024 \\ \hline
\end{tabular}
\label{Propeller_pro_ope}
\end{table}
\end{center}

For the propeller analysis, key non-dimensional operational parameters are the advance coefficient and cavitation number defined as
\begin{equation}\label{}
J=\frac{U_0}{nD}, \,\,\, \text{and} \,\,\, \sigma _n=\frac{p-p_{v}}{0.5\rho_l \left( nD \right) ^2},
\end{equation}
where $U_0$ is the incoming flow velocity (m/s), $n$ is propeller rotating rate (1/s), $D$ is propeller diameter (m), and $p_{v}$ is the vapor pressure (Pa). 
We monitor the propeller performance in the form of thrust and torque coefficients defined as
\begin{equation}\label{}
K_T=\frac{T}{\rho_l n^2D^4}, K_Q=\frac{Q}{\rho_l n^2D^5},
\end{equation}
where $T$ and $Q$ are the propeller thrust and torque respectively. The operating conditions for all the configurations considered in the present work are indicated in Table \ref{Propeller_pro_ope}.

The mesh dependency study is conducted via the 3-D simulation of the VP 1304 propeller without the surrounding duct. The corresponding thrust and torque coefficients at different mesh qualities are calculated and the associated results are summarized in Table \ref{thrust_com}.
It could be observed that the relative differences of each parameter between mesh 1 to mesh 2 are considerable, but all decrease to a value smaller as the mesh is refined to mesh 2 (fine) and mesh 3 (very fine).
To follow up, we use mesh 3 to calculate the present configuration and compare the dynamics coefficients between present results and other accessible experimental and numerical works \citep{PPTC_workshop, viitanen2019numerical}.  
The results for thrust and torque coefficients summarized in Table \ref{thrust_com} also indicate high conformance between this study and other results.
As a consequence, the strategy applied by mesh 3 is adopted in all the configurations of the present work to achieve the best balance of calculation time and accuracy.
Total cavitation volume is not compared herein because hub downstream cavitation is not considered and captured numerically herein owing to the meshing strategy.
Fig.~\ref{mesh1and2}a displays the overview of the mesh domain used in the present study, with the expanded/close-up views of mesh in the immediate vicinity of the propeller (and duct) shown in Fig.~\ref{mesh1and2}b.
Additionally, an intuitive diagram of the normalized cell size applied for the resolution of the tip vortex region, i.e., $\hat{x}_{tv}$ in Table \ref{thrust_com}, is shown in panel (iv) of Figure \ref{mesh1and2}b. $\hat{x}_{tv}$  (=${x}_{tv}/S$) represents the normalized cell size, where ${x}_{tv}$ is the maximum cell size in the tip vortex region.

\begin{center}
\begin{table*}
\caption{Thrust and torque coefficients for different mesh conditions. The overall mesh refinement level is adjusted by changing the base cell scale on the input/output patches. $\hat{x}_{tv}$  (=${x}_{tv}/S$) means the normalized cell size for the tip vortex region, where ${x}_{tv}$ is the maximum cell size in the tip vortex region.}
\centering
\begin{tabular}{  c  c  c  c  c  c  c  c  c  c  }
\hline
&  Total cells number & Rotor region & Stator region & $K_{T,rms}$ & 10$K_{Q,rms}$ & y$^+$ & $\hat{x}_{tv}$ \\ 
&     (millions)      &  (millions)  &  (millions)   &       &     &      &     \\
\hline
 Experiments \cite{PPTC_propeller} & & & & 0.374 & 0.9698 &  &  \\ 
 Other simulation \citep{viitanen2019numerical} & & & & 0.380 & 0.9680 &  &  \\ 
 Mesh 1 & 18.0 & 15.0 & 3.0 & 0.361 & 1.0398 & 92 & 0.004 \\ 
 Mesh 2 & 26.9 & 22.5 & 4.5 & 0.375 & 0.9710 & 40 & 0.006 \\ 
 Mesh 3 & 40.4 & 33.7 & 6.7 & 0.378 & 0.9648 & 19 & 0.009 \\ \hline
\end{tabular}
\label{thrust_com}
\end{table*}
\end{center}

\begin{figure}
  \centering
  \begin{subfigure}[b]{0.49\linewidth}
    \includegraphics[width=\linewidth]{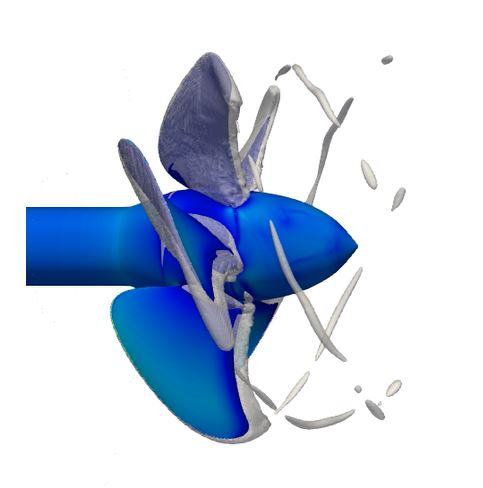}
    \caption{}
    \label{VP1304_ca_cavitation_des}
  \end{subfigure}
  \begin{subfigure}[b]{0.49\linewidth}
    \includegraphics[width=\linewidth]{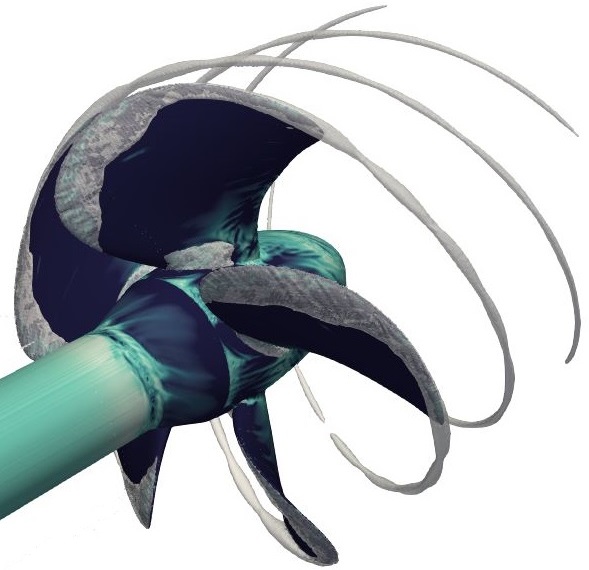}
    \caption{}
    \label{VP1304_ca_cavitation_les}
  \end{subfigure}
  
  \begin{subfigure}[b]{0.5\linewidth}
    \includegraphics[width=\linewidth]{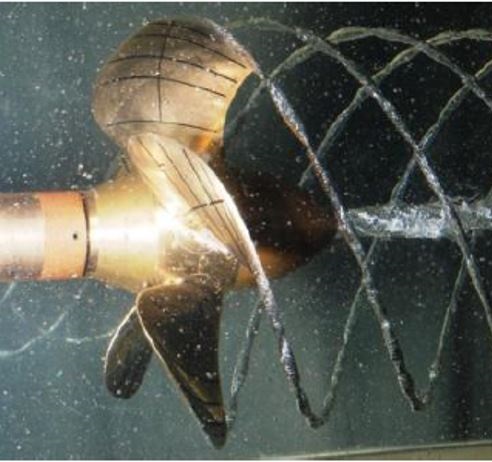}
    \caption{}
    \label{exp_VP1304_ca_cavitation}
  \end{subfigure}
  \caption{(a) Present numerical results obtained with DES model, (b) present numerical results obtained with LES model, and (c) experimental results \cite{PPTC_propeller}.}
  \label{VP1304_ca_cavitation_com}
\end{figure}

The resolution tip vortex cavitation, the key factor that affects the hydroacoustics generation, is highly sensitive to the applied numerical model, so we also compare the performance of $k-\omega$ SST Detached eddy simulation (DES) and LES models regarding the cavitation generation.
The cavitation contour with the volume fraction of 0.5 for DES and LES models are displayed in Fig. \ref{VP1304_ca_cavitation_com}a and b, respectively. 
In terms of the cavitation, it is apparent that the LES model exhibits better agreement with the experimental observation than the DES model and is thereby applied in present study.

\section{Flow dynamics}
\label{md_f}

\begin{figure}
\centering
\includegraphics[width=1.0\linewidth]{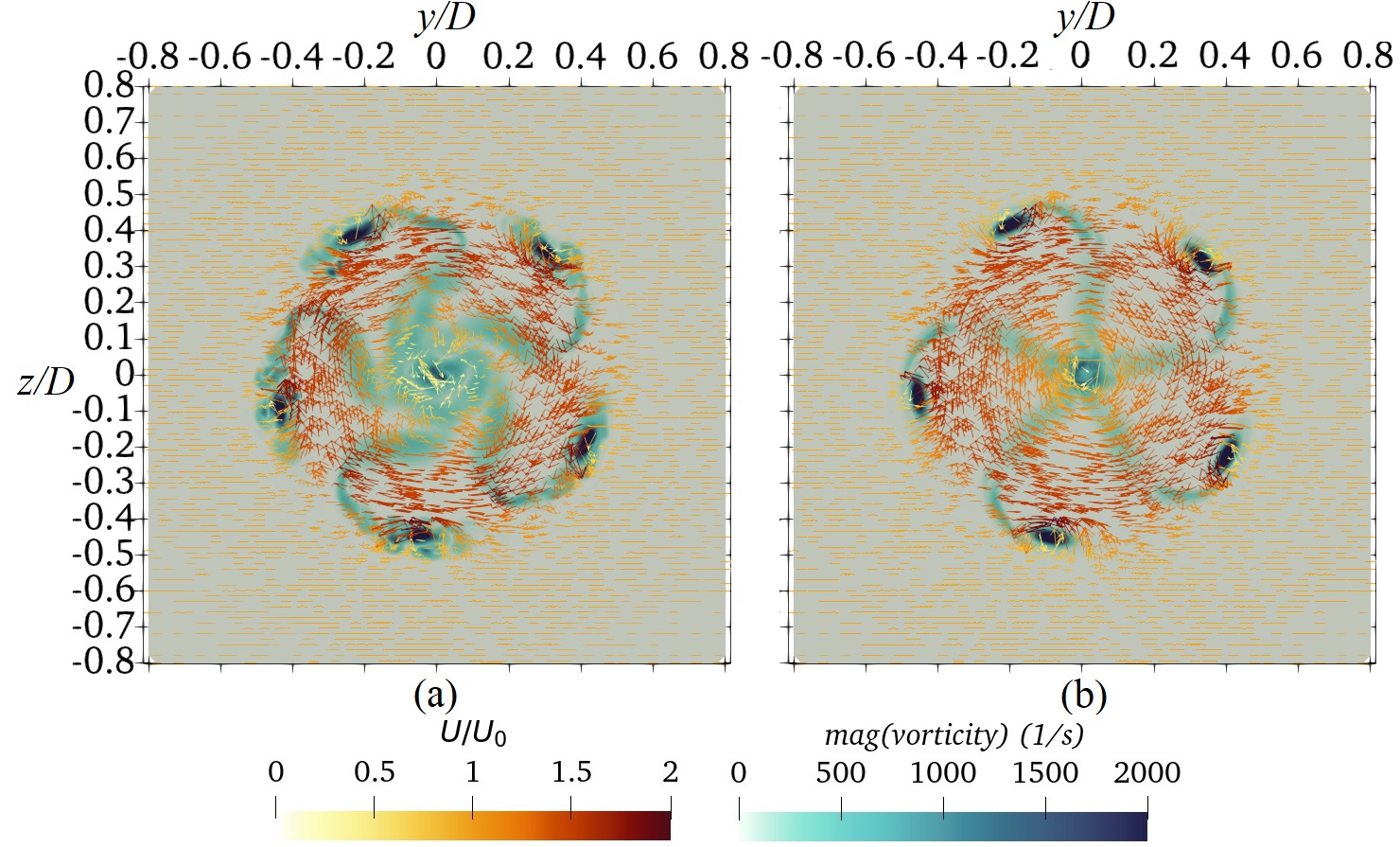}
\caption{Vorticity contour accompanied by streamlines in $y-z$ plane at $x=0.4D$ downstream for  (a) cavitating and (b) non-cavitating conditions.}\label{ca_noca_slice0.1m_U_vor}
\end{figure}


\begin{figure}
\centering
\includegraphics[width=1.0\linewidth]{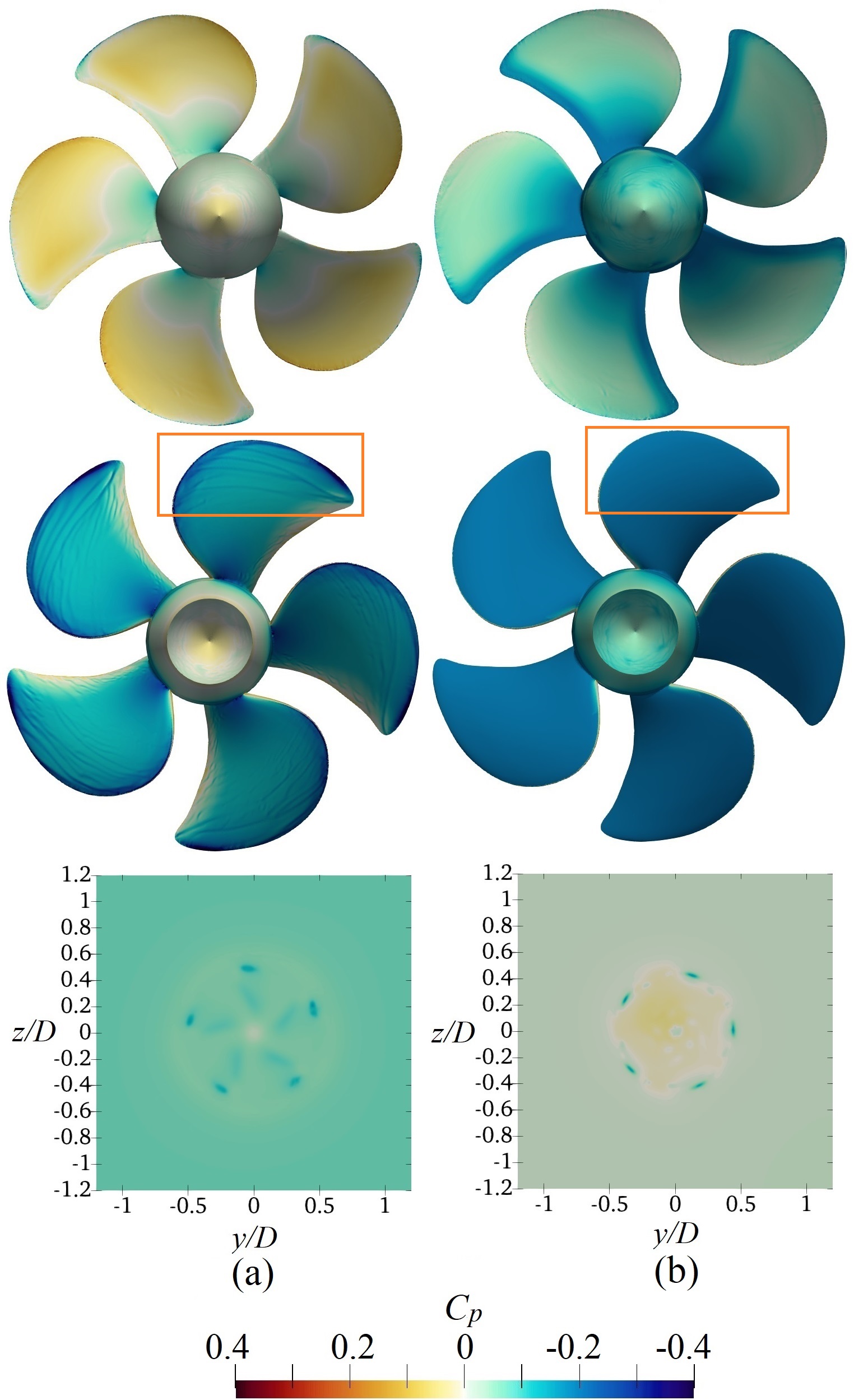}
\caption{Contour of pressure coefficients $C_p$ on the propeller surfaces and downstream $y-z$ planes (at $x=0.4D$) for (a) cavitating, and (b) non-cavitating conditions.}\label{ca_noca_surface_cp}
\end{figure}

In this section we focus on the hydrodynamic performance of the VP1304 propeller under cavitating and non-cavitating conditions.
To achieve non-cavitating conditions we increase the far-field  pressure thereby increasing the cavitation number, which is consistent with the experimental practices.
Figure \ref{ca_noca_slice0.1m_U_vor} shows the vorticity contour and velocity streamlines in the $y-z$ plane at $x=0.4D$ downstream of the propeller.
The length and direction of the small arrows represent the magnitude and direction of the flow velocity, respectively. 
Cavitating conditions result in localized complex patterns in the vortex distribution surrounding the tip vortices trajectory, which are caused by the cavity structure's effect on the flow field. This influences the flow direction in the surrounding fluid and may result in changes in the characteristics of noise sources.

Figure \ref{ca_noca_surface_cp} displays the distribution of the pressure coefficient $C_P$ on the propeller surfaces and the $y-z$ plane (at $x=0.4D$).
$C_p=(p-p_{ref})/(0.5\rho_l \left( n \pi D \right) ^2) $, and $p_{ref}$ is the reference pressure equal to $1 \times 10^5$ Pa.
The three panels on the left (Figure \ref{ca_noca_surface_cp}a) demonstrate flow characteristics for cavitating conditions. On the pressure side of the blades, there is a high-pressure region at the leading edge of the blade. Additionally, low-pressure zones are observed at the root and tip locations. This is consistent with the locations where cavities are observed in the experiments. Further, as seen in the $y-z$ planar section the pressure value along the tip vortex trajectory is small and corresponds to regions of tip vortex cavitation.

The three panels on the right (Figure \ref{ca_noca_surface_cp}b) demonstrate corresponding flow characteristics for non-cavitating conditions. The cavitation number for the non-cavitating case is set to a high value, and the resulting pressure coefficient is lower on both the suction and pressure sides when compared to the cavitating case.
A perusal of the suction side of the propeller shows that the pressure distribution on the blade surface is less uniform for the cavitating case as compared to the non-cavitation scenario, especially close to the tip of the blade (identified by the yellow box). This is expected to be associated with the sliding trajectory of the cavitation structure on the blade. This pressure disturbance enhances the energy of the noise source and affects the noise propagation.



\begin{figure}[h]
  \centering
  \begin{subfigure}{0.495\linewidth}
    \includegraphics[width=\linewidth]{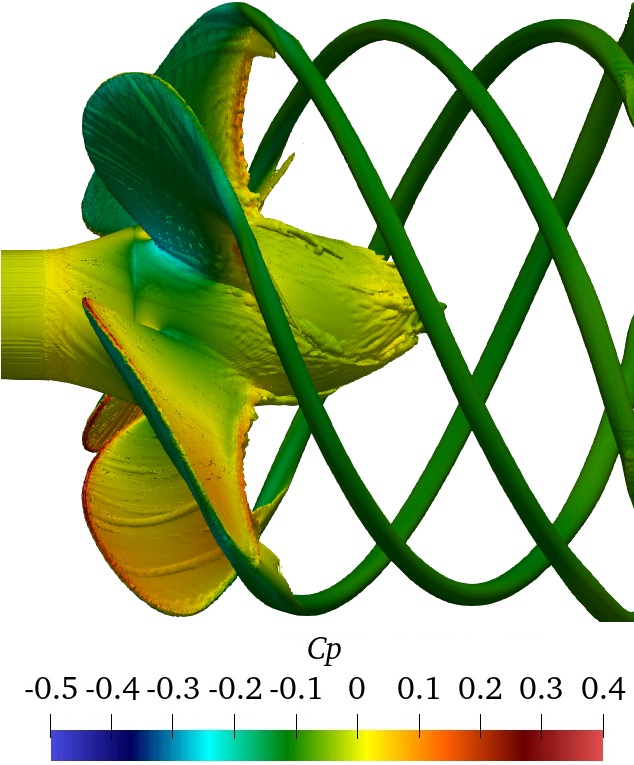}
    \caption{}
    \label{noca_vorticitycontour_pbar}
  \end{subfigure}
  \begin{subfigure}{0.495\linewidth}
    \includegraphics[width=\linewidth]{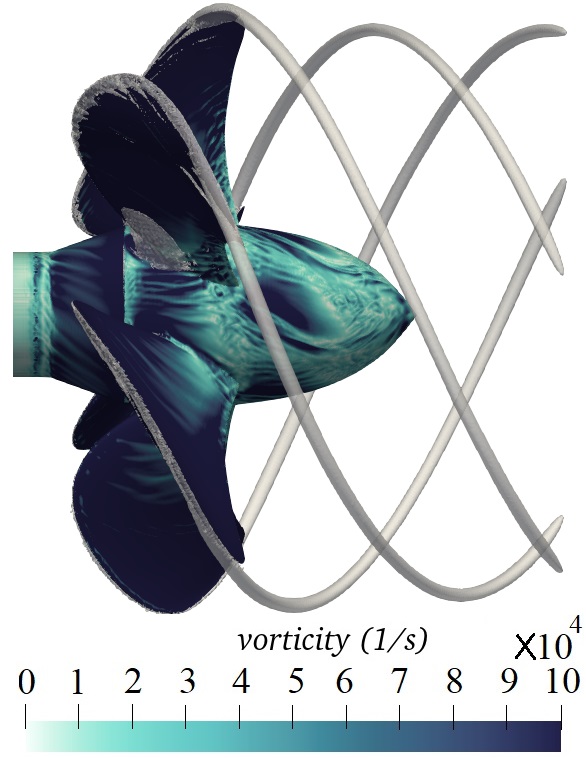}
    \caption{}
    \label{noca_Pvcontour_vorbar}
  \end{subfigure}
  
  \begin{subfigure}{0.495\linewidth}
    \includegraphics[width=\linewidth]{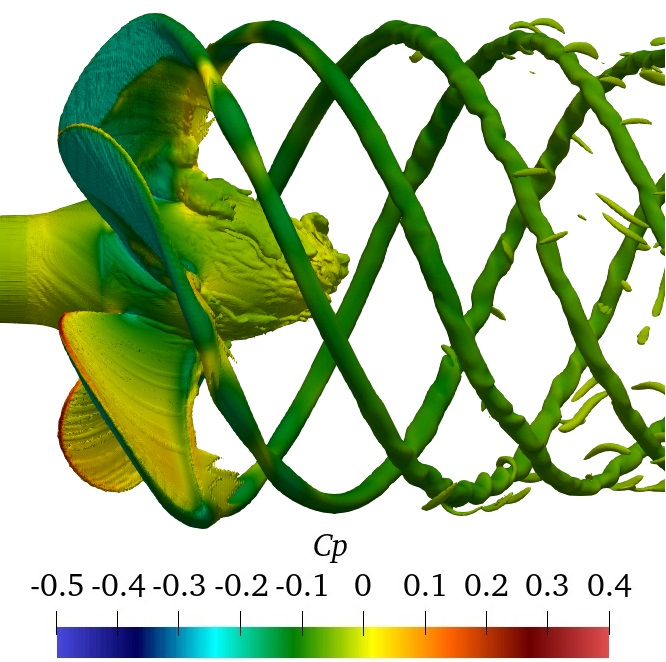}
    \caption{}
    \label{ca_vorticitycontour_Pbar}
  \end{subfigure}
  \begin{subfigure}{0.495\linewidth}
    \includegraphics[width=\linewidth]{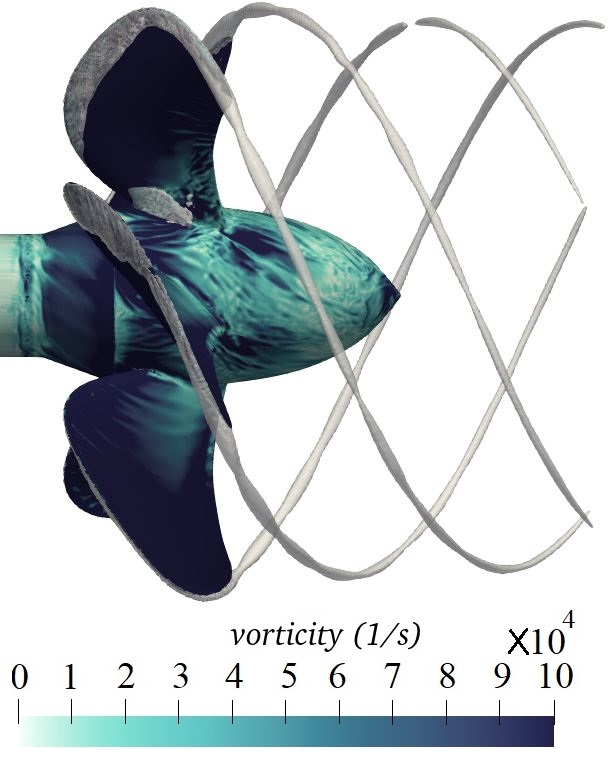}
    \caption{}
    \label{noca_Avcontour_pbar}
  \end{subfigure}  
  \caption{(a) vorticity isosurface colored by pressure coefficients, and (b)pressure isosurface colored by vorticity for non-cavitating condition; (c) Vorticity isosurface colored by pressure coefficients, and (d) cavitation contours (with volume fraction of 0.5) colored by vorticity for cavitating condition.}
  \label{VP1304_noca_ca_com}
\end{figure}

Figure \ref{VP1304_noca_ca_com} compares the wake structures of the VP1304 propeller in non-cavitating (top panels) and cavitating (bottom panels) conditions. 
Panel (a) shows the isosurface of vorticity (with value of 1500 1/s) colored by the pressure coefficient.
Clear tip vortices could be observed shedding from the tips of the rotating blades and extending farther downstream.
Consistent with experimental measurements, the cavitation number $\sigma _n$ here is 2.024 leading to the saturation pressure $p_v$ of 60,548 Pa (with a far-field pressure $p_{\infty}$ of 100,000 Pa).
In this case, we extract the iso-contour of pressure $p= 60,548 Pa$. The obtained pressure contour colored by the vorticity is shown in panel (b) of Fig. \ref{VP1304_noca_ca_com}. Pressure profiles are shed regularly from the blade tips and, in addition, the regions of low-pressure appear at the roots of the blades. This is consistent with experimental observations \cite{PPTC_propeller}.
In our further work, present numerically-obtained non-cavitating vortex behaviors will be compared with the available experimental results.

\begin{figure}
\centering
\includegraphics[width=1.0\linewidth]{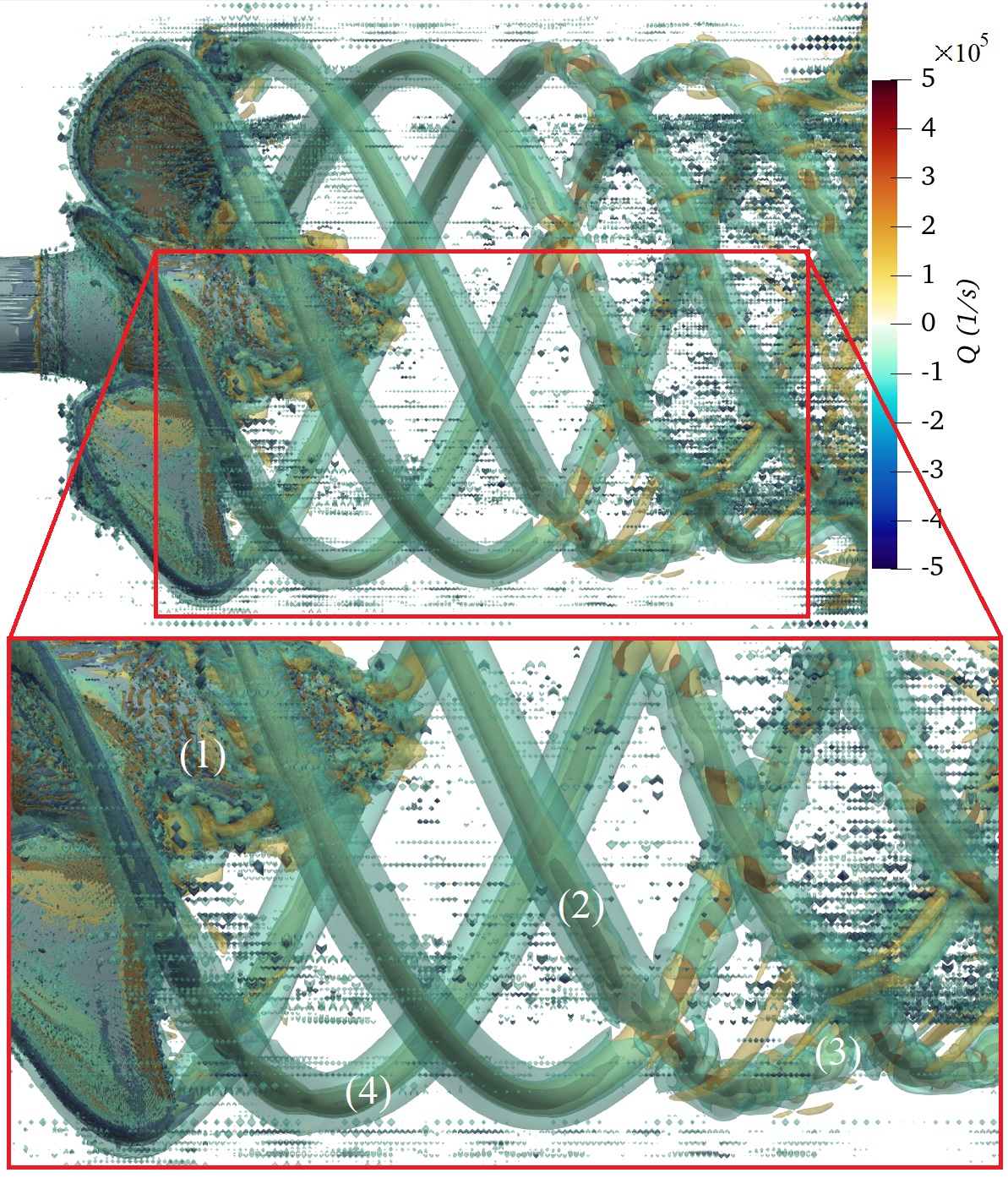}
\caption{Instantaneous isopleths of the second invariant $Q$ in the vicinity of the propeller for cavitating condition.}\label{ca_3DPropeller_Q_20240101}
\end{figure}

\begin{figure*}[h]
\centering
\includegraphics[width=0.8\linewidth]{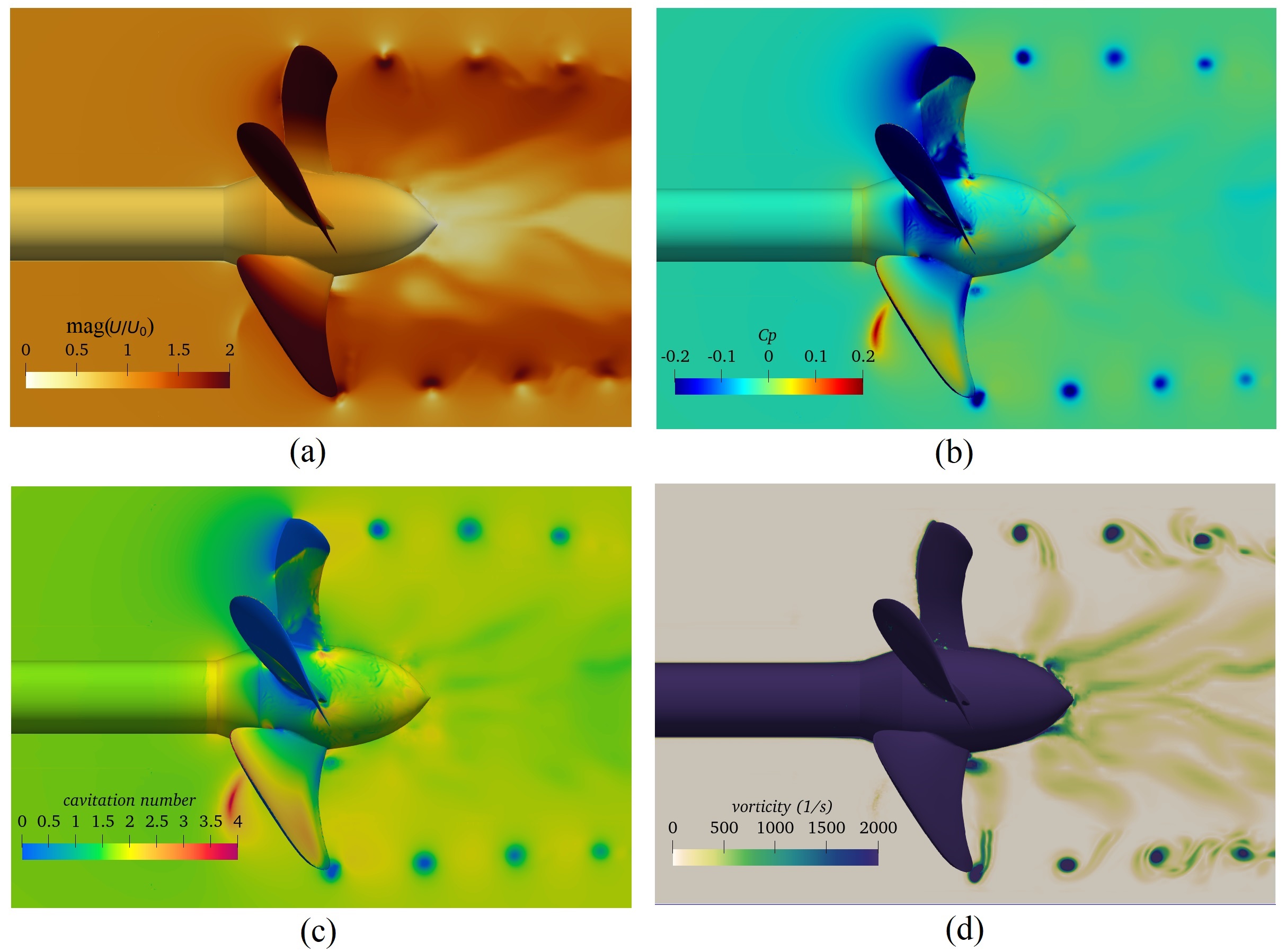}
\caption{Propeller surface accompanied by the $x-y$ planes, on which (a) normalized velocity magnitude mag$U/U_0$, (b) pressure coefficients $C_p$, (c) cavitation number $\sigma _n$, and (d) vorticity magnitude are depicted.}\label{ca_yzplane}
\end{figure*}

Figure \ref{VP1304_noca_ca_com} (Panel c) shows the propeller wake in the cavitating condition.  Initially, the vortices display a well-organized structure, maintaining distinct characteristics in the wake. However, the filaments gradually destabilize as they progress downstream. The tip vortices exhibit a dominant vortex and a secondary vortex, the latter dissipating swiftly along the extension. This observed behavior aligns with previous observations in both full propeller \cite{MUSCARI201365, ZHANG2019202} and also hydrofoil \cite{ye_wang_shao_2023}. 
The cavitation contour with a volume fraction of 0.5 is shown in panel (d) of Fig. \ref{VP1304_noca_ca_com}. 
By comparing panels (b) and (d), it can be expected that the intervention of the cavitation model allows the tip vortex to encompass the torsional mode and indicates a double-helical cavitation pattern \cite{ye_wang_shao_2023}, which is consistent with the experimental shot as shown in Fig. \ref{VP1304_ca_cavitation_com}(c).

To further analyze the transition behavior, an instantaneous visualization using isosurfaces of $Q$ is illustrated in Fig. \ref{ca_3DPropeller_Q_20240101}, where $Q$ is the second invariant of the velocity gradient tensor under a Galilean transformation.
It should be noted that $Q$ represents the difference between rotation rate and strain rate, and its positive value indicates that the vortex structure is rapidly forming at a correlated location. There are a large number of regions with positive and negative values of $Q$ on the pressure side of the blade (marked by white-color number \#1), indicating abundant vortex structures herein. In addition, it could be found that there is a clear vortex shedding behind the blade tip (as marked by white-color number \#2). Moreover, a clear torsional vortex feature caused by the double-helical cavitation structure could be observed at location \#4.
It is expected that the tip vortex (cavitation) and the surface sheeting vortex correspond to tonal and broadband components in the correlated noise source.
Furthermore, as demonstrated by the extension of the $Q$ contour trajectory from location \#2 to \#3, the tip vortices lose their coherence and start to break up into smaller structures. More specifically, the short-wave instability \cite{Widnall_1972} generates inside the vortex cores, and vortex oscillation is also observed to induce the second vortex structures. Further destabilization of those vortices leads to turbulence with small structures, bringing turbulence-induced noise sources and exhibiting broadband features in the spectrum.

Figure \ref{ca_yzplane} exhibits instantaneous contour of normalized magnitude velocity mag($U/U_0$), pressure coefficients $C_p$, cavitation number $\sigma _n$, and magnitude vorticity. Panel (a) indicates that the serrated high-speed flow pattern is generated after the rotating blades, accompanied by the low-pressure area formed at the tip vortex region (panel (b)). This is also indicated by panel (c) which shows the distribution of cavitation number. In panel (d), the vortex is presented to shed from the blade tip and hub center in the near wake region, and the thin trailing edges wake generated by the rotating blades. Consistent with previous observations, the tip vortex becomes unstable on extending further downstream.

\begin{figure}
\centering
\includegraphics[width=0.7\linewidth]{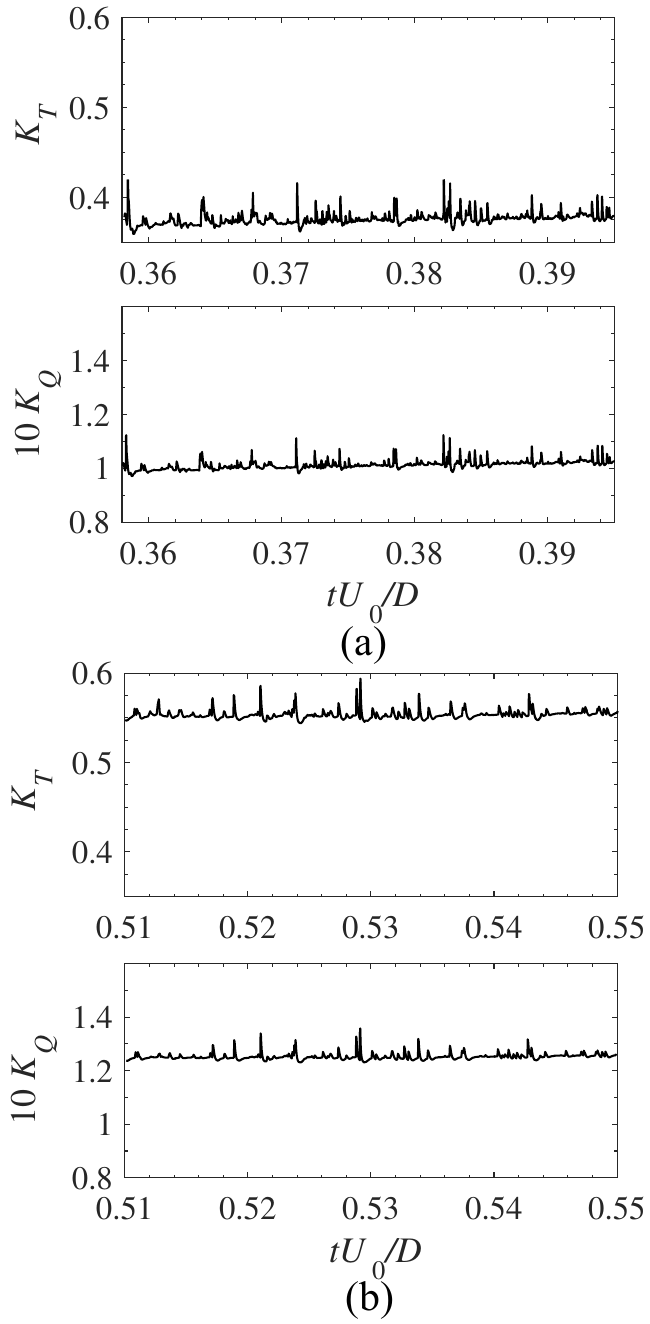}
\caption{Time histories of the propeller thrust coefficients $K_T$ and torque coefficients 10$K_Q$ for (a) cavitating and (b) non-cavitating conditions.}\label{noca_ca_kt_kq_time_20231225}
\end{figure}

Table \ref{threecase_KQ_com} compares the Root Mean Square (rms) value of the thrust and torque coefficients (i.e., $K_T$ and 10$K_Q$) between cavitating and non-cavitating conditions for the VP1304 propeller at specified operation conditions.
We find that both $K_{T,rms}$ and 10$K_{Q,rms}$ decrease in the cavitating scenario, owing to change in the pressure distribution on the blade surfaces.
Figure \ref{noca_ca_kt_kq_time_20231225} displays the time histories of the propeller thrust coefficients $K_T$ and torque coefficients $K_Q$ for cavitating and non-cavitating conditions, in which the top and bottom panels show the cavitating and non-cavitating cases, respectively. 
We note that the time variations of $K_T$ and $K_Q$ reach periodic equilibrium in both the cavitating and non-cavitating cases.

\begin{center}   
\begin{table}[!h]
\caption{Comparison for $K_{T,rms}$ and 10$K_{Q,rms}$ between cavitating and non-cavitating conditions.}
\centering
\begin{tabular}{  c  c  c  c  c  c  c  c  c  c  }
\hline
&  $K_{T,rms}$ & 10$K_{Q,rms}$  \\
\hline
 Non-cavitating &  0.542 & 1.2421 \\ 
 Cavitating &  0.378 & 0.9648  \\ \hline
\end{tabular}
\label{threecase_KQ_com}
\end{table}
\end{center}

\begin{figure}
\centering
\includegraphics[width=1.0\linewidth]{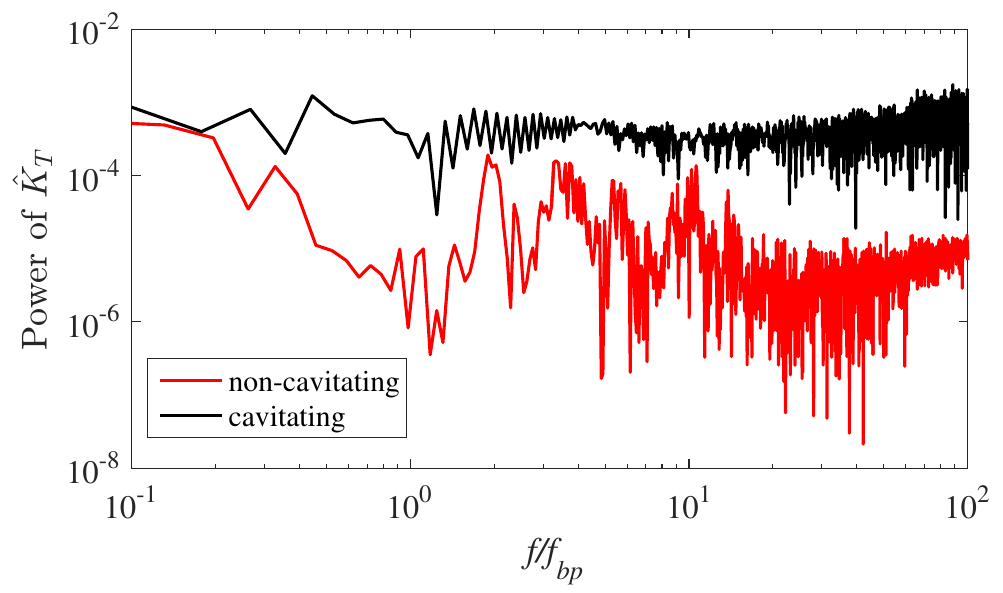}
\caption{Power spectrum of fluctuating component $\hat{K}_T$ of the thrust coefficient $K_T$ for cavitating and non-cavitating conditions.}\label{noca_ca_KT_spectrum}
\end{figure}

Since the time variations in both $K_T$ and $K_Q$ are subject to the total pressure fluctuations on the propeller surface, the fluctuations in both thereby remain consistent. This is also confirmed by a careful comparison of the time-histories of $K_T$ and $K_Q$. In this case, we further extracted the fluctuating components in the variation of $K_T$  (omitting the $K_{T,rms}$) for the non-cavitating and cavitating cases, and plotted their spectrum in Fig. \ref{noca_ca_KT_spectrum}. 
The data in one full rotation cycle (i.e., 4 seconds) with the non-dimensional time step ($tU_0/D$) of 2.55$\times 10^{-5}$ are obtained herein for spectrum analysis.
The frequency is normalized by the blade passing frequency $f_{bp}$.

As displayed in Fig. \ref{noca_ca_kt_kq_time_20231225}, although the mean value of $K_T$ is higher for the non-cavitation case, the energy of its fluctuation components $\hat{K}_T$ is weaker. This is also implied in Fig. \ref{noca_ca_KT_spectrum} where the variation in the power spectrum for the cavitating case is stronger than that of the non-cavitating case concerning the intensity of peaks at low frequencies and broadband components at high frequencies. 
This suggests that the generation of cavitation enhances the pressure fluctuations on the propeller surface as a whole. In addition, the occurrence of peaks in the low-frequency range increase in the cavitating case, which is expected to be correlated to the generation of cavitation structures at the tip of the blade and their detachment/shedding from the tip. Specifically, the appearance of tip vortex cavitation leads to the amplification of the loading (or monopole) component underlying the noise mechanism. In addition, the enhancement of broadband frequencies under the cavitating condition is associated with more complex flow field chaos at the surface and also the generation of abundant tiny bubbles.


\section{Hydroacoustics}
\label{aa_section}

\begin{figure}
\centering
\includegraphics[width=0.8\linewidth]{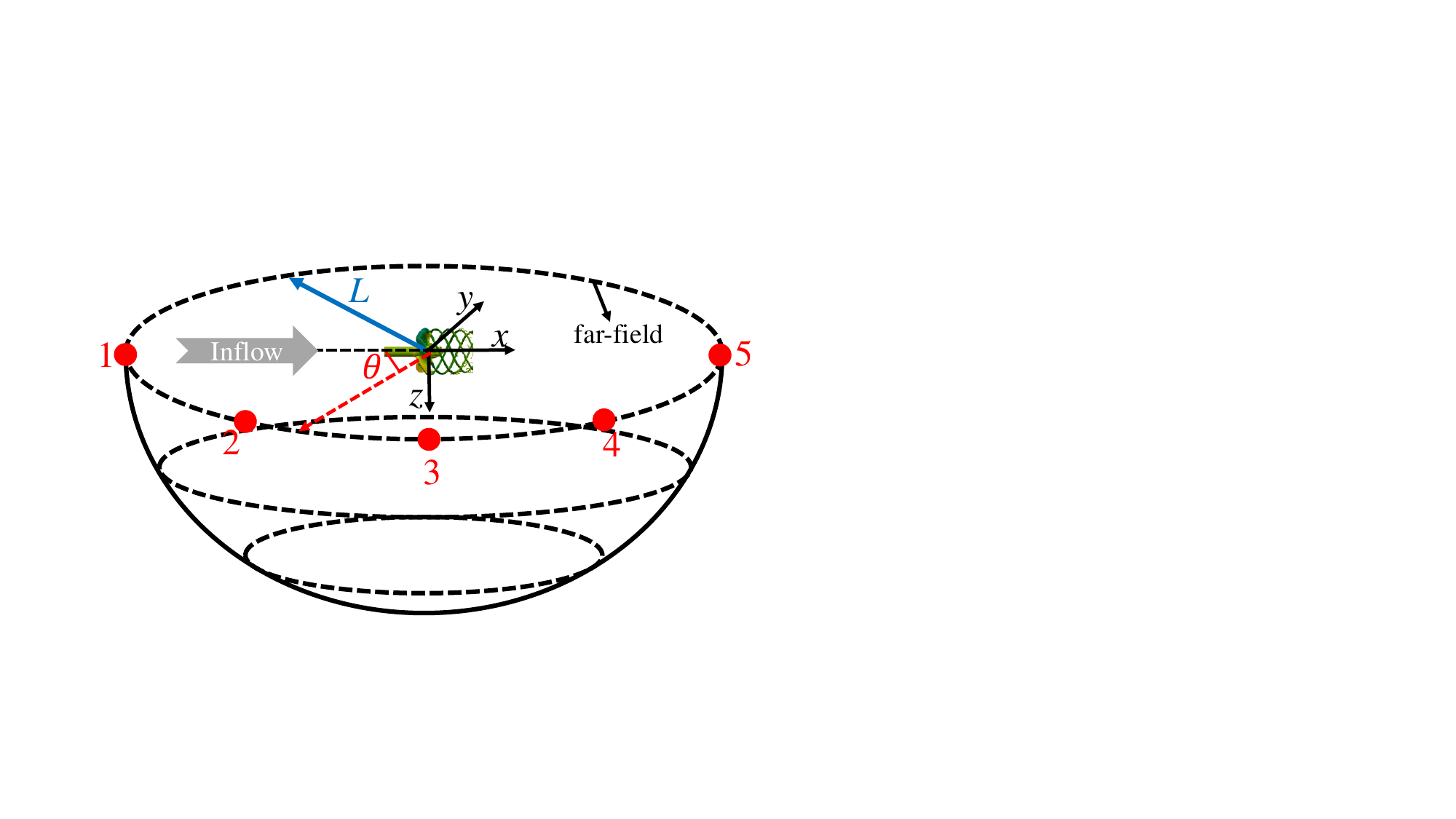}
\caption{The location of the microphones.}\label{acoustics_locations}
\end{figure}

The following discussion on aeroacoustics will be presented based on the configuration displayed in Figure \ref{acoustics_locations}. Five acoustics pressure monitors are located in the angular $\theta$ of 0$^{\circ}$, 45$^{\circ}$, 90$^{\circ}$, 135$^{\circ}$, and 180$^{\circ}$ with distance $L$ of 100 m and 1000 m from the rotor center. 
One permeable FW-H control surface is applied to record the real-time fluctuation of acoustics pressure in the near- and far-fields.

\begin{figure}
  \centering
  \begin{subfigure}[b]{1.0\linewidth}
    \includegraphics[width=\linewidth]{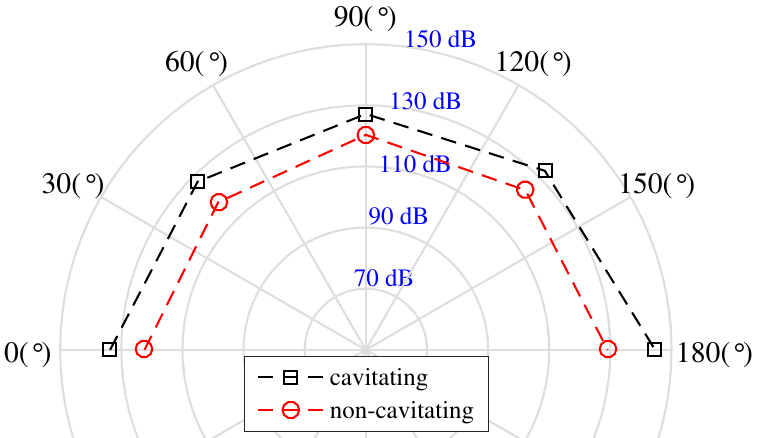}
    \caption{}
    \label{ca_noca_com_Re100m_SPL}
  \end{subfigure}
  
  \begin{subfigure}[b]{1.0\linewidth}
    \includegraphics[width=\linewidth]{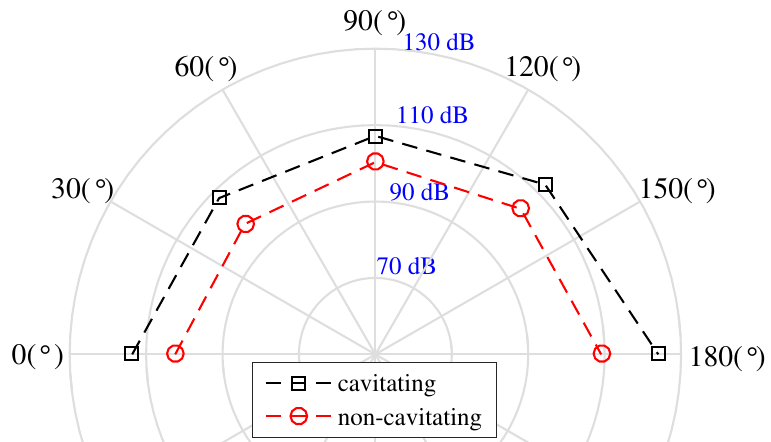}
    \caption{}
    \label{ca_noca_com_Re1000m_SPL}
  \end{subfigure}  
  \caption{SPL directivity at orientations $\theta$ of 0$^{\circ}$, 45$^{\circ}$, 90$^{\circ}$, 135$^{\circ}$, and 180$^{\circ}$ with distances $L$ of (a) 100 m and (b) 1000 m for cavitating and non-cavitating conditions.}
  \label{ca_noca_com_Re100_1000m_SPL}
\end{figure}

The value of Sound pressure level (SPL) on a circumference (with a radius of 100 $m$ from the propeller center) at angles ranging from 0$^{\circ}$ and 180$^{\circ}$ are identified in Fig.~\ref{ca_noca_com_Re100m_SPL}, in which cavitating and non-cavitating conditions are compared.
SPL (dB) = 20log($p'_{rms}$ /$p_{h,ref}$), where $p'$ is the acoustics pressure fluctuation and $p_{h,ref}$ is the hydroacoustics reference pressure of $1\times 10^{-6}$ Pa.
In terms of the SPL directivity with a radius of 100 m for the cavitating case (cf. with panel (a)), the acoustics energy is highest in the downstream direction (180$^{\circ}$), where it could reach 150 dB, and lowest at the side direction (90$^{\circ}$), where it is about 125 dB.
This phenomenon is also observed in the non-cavitating case. It is expected that this behavior is correlated to the extension of the propeller wake in the streamlined direction.

Overall, as the distance increases from 100 m to 1000 m, the SPLs of cavitating and non-cavitating cases decrease by about 25dB and 20dB, respectively, on average. This is owing to that the fluctuating energy of sound pressure is gradually dissipated by the medium viscosity while propagating to the far field.
Furthermore, it is obvious from both panels (a) and (b) that the generation of cavitating structures will significantly increase the URN power. In more detail, a perusal of the figures indicates that when cavitation is generated, the increment/increase of SPL is essentially equal in all directions, indicating that the noise source induced by cavitation has an omnidirectional pattern and demonstrates a monopole source mechanism.
However, the SPL increase is slightly larger in the downstream direction compared to those of the upstream and side directions.
Based on the analysis of Figs. \ref{ca_noca_slice0.1m_U_vor} and \ref{ca_3DPropeller_Q_20240101}, cavitation generation leads to a more irregular vortex structure in the wake region and those vortices are more prone to fragment, which nevertheless brings about noise enhancement in the downstream direction.

\begin{figure}
\centering
\includegraphics[width=1.0\linewidth]{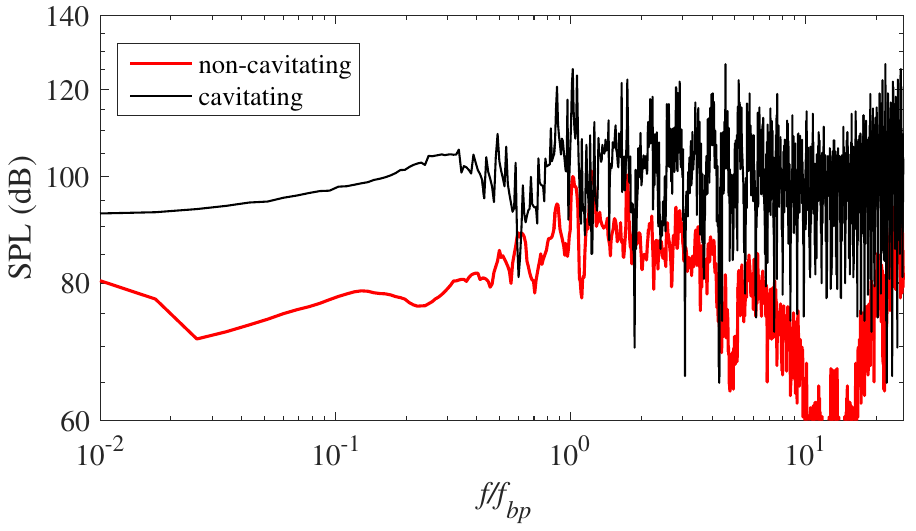}
\caption{The comparison of SPL spectrum between cavitating and non-cavitating conditions at directivity angles of 90$^{\circ}$ with a distance of 1000 m.}\label{noca_ca_SPL_spectrum_R100_180deg_20240105}
\end{figure}

Figure \ref{noca_ca_SPL_spectrum_R100_180deg_20240105} depicts the spectrum of SPL obtained at the monitor with directivity angles of 90$^{\circ}$ and a distance of 1000 m, and the cavitating and non-cavitating conditions are compared.
The acoustics energy exhibited by the data for the cavitating condition is generally stronger across the overall frequency range than that of the non-cavitation condition. This implies that the acoustics pressure obtained with cavitation generation is more energetic regarding no matter the low-frequency tonal or high-frequency broadband components. This is consistent with the features exhibited by the thrust spectrum in Fig. \ref{noca_ca_KT_spectrum}.
Specifically, several intense peaks are located at $f/f_{bp}$ = 1, 2, and 3, meaning that blade passing behaviors and accompanied shedding behaviors of tip vortex (cavitation) are closely correlated to the tonal noise components.
According to the FW-H theory, noise sources of monopoles (or loading component), dipoles (thickness component), and quadrupoles (turbulence-induced component) together constitute the propagation of hydrodynamic noise. Furthermore, the turbulence-induced noise sources are due to wake dynamics unsteadiness, which is related to the complicated vortex patterns. This is qualitatively shown in Figs. \ref{VP1304_noca_ca_com} and \ref{ca_3DPropeller_Q_20240101}.
Furthermore, the reason for thickness noise is the displacement of the fluid pushed by the rotating of the propeller body.
Regarding the noise sources, the loading noise exhibiting monopole feature originates from the accelerating force distribution on the propeller surface and in particular from the influence of the load distribution. The cavitation structure and associated pressure distribution variation on the blade surface will therefore be primarily responsible for the enhancement of this component. 




\section{Conclusion}
This study presented a comprehensive numerical investigation into the flow dynamics, cavitation patterns, and hydroacoustic propagation associated with a five-blade full propeller and further explored the potential mechanism underlying the propeller noise radiation. Our methodology includes a standard dynamic Large Eddy Simulation (LES) model, the Schnerr-Sauer model for cavitation, and the Ffowcs-Williams-Hawkings (FW-H) method. Both cavitating and non-cavitating conditions are considered in the present work. 
Our modeling results, when compared to experimental data, validate the accuracy and reliability of our computational approach. In the cavitating configuration, the formation of cavitation occurs at the root and tip of the propeller blades, with the observation of a distinct double-helical pattern in the tip vortex cavitation. The trajectory analysis of the tip vortex showcases its evolution from discernible filaments to destabilization and eventual vortex breakdown.
Notably, the pressure fluctuations experienced on the propeller surface during cavitating conditions exhibit more pronounced variability compared to non-cavitating conditions. This intensified fluctuation amplifies the energy emitted by the noise source. Our comparative assessment of the sound pressure level (SPL) directionality confirms that the hydroacoustic power generated during cavitating scenarios surpasses that of non-cavitating instances in all directions, particularly evident in the downstream direction with an increase in SPL of up to 20 dB.
The generation of sheet cavitation and tip vortex cavitation leads to the enhancement of monopole (loading) noise source, and the associated collapsing of vortices and bubbles in the wake contributes to broadband noise components. Additionally, the tonal components, at frequencies corresponding to harmonics of blade passing frequency, intensify owing to the close correlation between the structure of tip vortex cavitation and the blade passing.
The interaction of the cavitating wake with the propeller nozzle and/or ship hull will be considered in future work.

\section*{Acknowledgment} 
The present study is supported by Mitacs, Transport Canada and Clear Seas through a Quiet-Vessel Initiative (QVI) program. 
The current study was partly supported by Robert Allan Ltd. through an Innovative Solutions Canada Phase 2 grant from Transport Canada's Innovation Centre.
We also acknowledge the computing facilities of Advanced Research Computing (ARC) at the University of British Columbia, Shared Hierarchical Academic Research Computing  (\href{http://www.sharcnet.ca}{SHARCNET}), and Compute/Calcul Canada.

\bibliographystyle{asmeconf}  
\bibliography{manuscripts_fullpropeller}

\begin{thebibliography}{10}
\newcommand{\enquote}[1]{``#1''}
\providecommand{\url}[1]{\texttt{#1}}
\providecommand{\urlprefix}{URL }
\expandafter\ifx\csname urlstyle\endcsname\relax
  \providecommand{\doi}[1]{DOI \discretionary{}{}{}#1}\else
  \providecommand{\doi}{DOI \discretionary{}{}{}\begingroup \urlstyle{rm}\Url}\fi
\providecommand{\eprint}[2][]{\urlprefix\url{#1#2}}

\bibitem{SMITH2022112863}
Smith, T.~A. and Rigby, J.
\newblock \enquote{Underwater radiated noise from marine vessels: A review of noise reduction methods and technology.}
\newblock \textit{Ocean Engineering} Vol. 266 (2022): p. 112863.

\bibitem{Savas2021}
S.~Sezen, M.~Atlar and Fitzsimmons, P.
\newblock \enquote{Prediction of cavitating propeller underwater radiated noise using RANS \& DES-based hybrid method.}
\newblock \textit{Ships and Offshore Structures} Vol.~16 No. sup1 (2021): pp. 93--105.

\bibitem{Asnaghi2020}
Asnaghi, A., Svennberg, U., Gustafsson, R. and Bensow, R.~E.
\newblock \enquote{{Investigations of tip vortex mitigation by using roughness}.}
\newblock \textit{Physics of Fluids} Vol.~32 No.~6 (2020): p. 065111.

\bibitem{A-Man2023}
Zhang, A.-M., Li, S.-M., Cui, P., Li, S. and Liu, Y.-L.
\newblock \enquote{{A unified theory for bubble dynamics}.}
\newblock \textit{Physics of Fluids} Vol.~35 No.~3 (2023): p. 033323.

\bibitem{Arveson2000}
Arveson, P. and Vendittis, D.~J.
\newblock \enquote{Radiated noise characteristics of a modern cargo ship.}
\newblock \textit{The Journal of the Acoustical Society of America} Vol. 107 (2000): pp. 118--29.

\bibitem{jmse9070778}
Kimmerl, J., Mertes, P. and Abdel-Maksoud, M.
\newblock \enquote{Application of Large Eddy Simulation to Predict Underwater Noise of Marine Propulsors. Part 2: Noise Generation.}
\newblock \textit{Journal of Marine Science and Engineering} Vol.~9 No.~7 (2021): p. 778.

\bibitem{seol2002prediction}
Seol, H., Jung, B., Suh, J.-C. and Lee, S.
\newblock \enquote{Prediction of non-cavitating underwater propeller noise.}
\newblock \textit{Journal of sound and Vibration} Vol. 257 No.~1 (2002): pp. 131--156.

\bibitem{VariantsoftheFfowcsWilliams}
Spalart, P. and Shur, M.
\newblock \enquote{Variants of the {Ffowcs Williams-Hawkings} equation and their coupling with simulations of hot jets.}
\newblock \textit{International Journal of Aeroacoustics} Vol.~8 (2009): pp. 477--491.

\bibitem{OntheuseoftheFfowcsWilliams}
Mendez, S., Shoeybi, M., Lele, S. and Moin, P.
\newblock \enquote{On the use of the {Ffowcs Williams-Hawkings} equation to predict far-field jet noise from large-eddy simulations.}
\newblock \textit{International Journal of Aeroacoustics} Vol.~12 (2013): pp. 1--20.

\bibitem{FILIOS20071497}
Filios, A.~E., Tachos, N.~S., Fragias, A.~P. and Margaris, D.~P.
\newblock \enquote{Broadband noise radiation analysis for an {HAWT} rotor.}
\newblock \textit{Renewable Energy} Vol.~32 No.~9 (2007): pp. 1497--1510.

\bibitem{posa_JFM_2022}
Posa, A., Broglia, R., Felli, M., Cianferra, M. and Armenio, V.
\newblock \enquote{Hydroacoustic analysis of a marine propeller using large-eddy simulation and acoustic analogy.}
\newblock \textit{Journal of Fluid Mechanics} Vol. 947 (2022): p. A46.

\bibitem{Shakeel2020}
Ahmed, Shakeel.
\newblock \enquote{On the Noise Generated by a Ship Propeller.}
\newblock Ph.D. Thesis, University of New South Wales.
\newblock 2020.

\bibitem{Posa2022}
Posa, A., Broglia, R. and Felli, M.
\newblock \enquote{{Acoustic signature of a propeller operating upstream of a hydrofoil}.}
\newblock \textit{Physics of Fluids} Vol.~34 No.~6 (2022): p. 065132.

\bibitem{yangzhou_wu_ma_huang_2023}
Yangzhou, J., Wu, J., Ma, Z. and Huang, X.
\newblock \enquote{Aeroacoustic sources analysis of wake-ingesting propeller noise.}
\newblock \textit{Journal of Fluid Mechanics} Vol. 962 (2023): p. A29.

\bibitem{ZHANG2019202}
Zhang, Q. and Jaiman, R.~K.
\newblock \enquote{Numerical analysis on the wake dynamics of a ducted propeller.}
\newblock \textit{Ocean Engineering} Vol. 171 (2019): pp. 202--224.

\bibitem{HIEKE2022112131}
Hieke, M., Sultani, H., Witte, M., {von Estorff}, O. and Wurm, F.-H.
\newblock \enquote{A workflow for hydroacoustic source analyses based on a scale-resolving flow simulation of a hubless propeller.}
\newblock \textit{Ocean Engineering} Vol. 261 (2022): p. 112131.

\bibitem{jmse10030378}
Chen, M., Liu, J., Si, Q., Liang, Y., Jin, Z. and Yuan, J.
\newblock \enquote{Investigation into the Hydrodynamic Noise Characteristics of Electric Ducted Propeller.}
\newblock \textit{Journal of Marine Science and Engineering} Vol.~10 No.~3 (2022).

\bibitem{PPTC_propeller}
\enquote{{PPTC smp’11 Workshop. [Online]. Available: https://www.sva-potsdam.de/en/pptc-smp11-workshop/}.}  .

\bibitem{Geese2022}
J.~Geese, M. Nadler M. A.-Maksoud, J.~Kimmerl.
\newblock \enquote{Acoustic Emissions and Cavitation Effects on Model Scale Propellers Using a Transition Model.}
\newblock \textit{Seventh International Symposium on Marine Propulsors smp’22, Wuxi, China, October 2022}: pp. 4--1--3. 2022.

\bibitem{VIITANEN2022112596}
Viitanen, V., Sipilä, T., Sánchez-Caja, A. and Siikonen, T.
\newblock \enquote{CFD predictions of unsteady cavitation for a marine propeller in oblique inflow.}
\newblock \textit{Ocean Engineering} Vol. 266 (2022): p. 112596.

\bibitem{Marta2019}
M.~Cianferra, V.~Armenio, A.~Petronio.
\newblock \enquote{Numerical prediction of ship propeller noise through acoustic analogy.}
\newblock \textit{Sixth International Symposium on Marine Propulsors smp’19, Rome, Italy, May 2019}: pp. WB3--3. 2019.

\bibitem{LIDTKE2022111176}
Artur, K.~L., Lloyd, T., Lafeber, F.~H. and Bosschers, J.
\newblock \enquote{Predicting cavitating propeller noise in off-design conditions using scale-resolving CFD simulations.}
\newblock \textit{Ocean Engineering} Vol. 254 (2022): p. 111176.

\bibitem{LESmodel}
Kim, W.-W. and Menon, S.
\newblock \enquote{A new dynamic one-equation subgrid-scale model for large eddy simulations.}
\newblock \textit{33rd Aerospace Sciences Meeting and Exhibit 09 January 1995 - 12 January 1995, Reno, NV, U.S.A.}

\bibitem{sauer2001development}
Sauer, J.
\newblock \enquote{Development of a new cavitation model based on bubble dynamics.}
\newblock \textit{ZAMM-Journal of Applied Mathematics and Mechanics/Zeitschrift f{\"u}r Angewandte Mathematik und Mechanik} Vol.~81 No. S3 S3 (2001): pp. 561--562.

\bibitem{asnaghi2017improvement}
Asnaghi, A, Feymark, A and Bensow, RE.
\newblock \enquote{Improvement of cavitation mass transfer modeling based on local flow properties.}
\newblock \textit{International Journal of Multiphase Flow} Vol.~93 (2017): pp. 142--157.

\bibitem{kashyap2023unsteady}
Kashyap, Suraj~R and Jaiman, Rajeev~K.
\newblock \enquote{Unsteady cavitation dynamics and frequency lock-in of a freely vibrating hydrofoil at high Reynolds number.}
\newblock \textit{International Journal of Multiphase Flow} Vol. 158 (2023): p. 104276.

\bibitem{CHENG2023652}
Cheng, Z., McConkey, R., Yee, E. and Lien, F.-S.
\newblock \enquote{Numerical investigation of noise suppression and amplification in forced oscillations of single and tandem cylinders in high Reynolds number turbulent flows.}
\newblock \textit{Applied Mathematical Modelling} Vol. 117 (2023): pp. 652--686.

\bibitem{farassat1988uses}
Farassat, F. and Brentner, K.~S.
\newblock \enquote{The uses and abuses of the acoustic analogy in helicopter rotor noise prediction.}
\newblock \textit{Journal of the American Helicopter Society} Vol.~33 No.~1 (1988): pp. 29--36.

\bibitem{wang2006computational}
Wang, M., Freund, J.~B. and Lele, S.~K.
\newblock \enquote{Computational prediction of flow-generated sound.}
\newblock \textit{Annual Review of Fluid Mechanics} Vol.~38 (2006): pp. 483--512.

\bibitem{farassat2007derivation}
Farassat, F.
\newblock \enquote{{Derivation of Formulations 1 and 1A of Farassat}.}
\newblock \textit{NASA/TM-2007–214853, (2007)}  .

\bibitem{farassat1981linear}
Farassat, F.
\newblock \enquote{Linear acoustic formulas for calculation of rotating blade noise.}
\newblock \textit{AIAA Journal} Vol.~19 No.~9 (1981): pp. 1122--1130.

\bibitem{farassat1975theory}
Farassat, F.
\newblock \textit{Theory of Noise Generation From Moving Bodies With an Application to Helicopter Rotors}.
\newblock NASA Langley Technical Report Server (1975).

\bibitem{farassat1998acoustic}
Farassat, F. and Brentner, K.~S.
\newblock \enquote{The acoustic analogy and the prediction of the noise of rotating blades.}
\newblock \textit{Theoretical and Computational Fluid Dynamics} Vol.~10 No. 1-4 (1998): pp. 155--170.

\bibitem{EPIKHIN2015150}
Andrey, E., Ilya, E., Matvey, K., Michael, K. and Sergei, S.
\newblock \enquote{{Development of a Dynamic Library for Computational Aeroacoustics Applications Using the OpenFOAM Open Source Package}.}
\newblock \textit{Procedia Computer Science} Vol.~66 (2015): pp. 150--157.
\newblock 4th International Young Scientist Conference on Computational Science.

\bibitem{openfoamv2006}
\enquote{{OpenCFD Ltd., OpenFOAM: User Guide v2006, 2019. [Online]. Available: https://www.openfoam.com/documentation/guides/latest/doc/.}}  .

\bibitem{PPTC_workshop}
Heinke, H.~J.
\newblock \enquote{{Potsdam Propeller Test Case (PPTC). Cavitation Tests with the Model Propeller VP1304}.}
\newblock \textit{SVA Potsdam Model Basin Report No.3753}.

\bibitem{viitanen2019numerical}
Viitanen, V., Siikonen, T. and S{\'a}nchez-Caja, A.
\newblock \enquote{Numerical Viscous Flow Simulations of Cavitating Propeller Flows at Different Reynolds Numbers.}
\newblock \textit{Proceedings of the Sixth International Symposium on Marine Propulsors, Rome, Italy}: pp. 26--30. 2019.

\bibitem{MUSCARI201365}
Muscari, R., {Di Mascio}, A. and Verzicco, R.
\newblock \enquote{Modeling of vortex dynamics in the wake of a marine propeller.}
\newblock \textit{Computers \& Fluids} Vol.~73 (2013): pp. 65--79.

\bibitem{ye_wang_shao_2023}
Ye, Q., Wang, Y. and Shao, X.
\newblock \enquote{Dynamics of cavitating tip vortex.}
\newblock \textit{Journal of Fluid Mechanics} Vol. 967 (2023): p. A30.

\bibitem{Widnall_1972}
Widnall, S.~E.
\newblock \enquote{The stability of a helical vortex filament.}
\newblock \textit{Journal of Fluid Mechanics} Vol.~54 No.~4 (1972): p. 641–663.

\end{thebibliography}

\appendix   



\selectlanguage{english} 

\end{document}